\documentstyle[12pt]{article}
\begin{document}
\def\as{\alpha_s}
\def\ee{e^+e^-}
\def\qq{q \bar{q}}
\def\lmsb{\Lambda_{\overline{\rm MS}}}
\def\ycut{y_{\rm cut}}
\def\yeff{y_{\rm eff}}
%%%%%%%%%%%%%%%%%%%%%%%%%%%%%
\def\JADE{1}
\def\bethke{2}
\def\kn{3}
\def\kl{4}
\def\bsi{5}
\def\bsii{6}
\def\yuri{7}
\def\LUCLUS{8}
\def\catani{9}
\def\ert{10}
\def\pms{11}
\def\fac{12}
\def\opaljet{13}
\def\jetset{8}
\def\opalshape{14}
\def\herwig{15}
%\def\nlo{}
%\def\cs{}
%%%%%%%%%%%%%%%%%%%%%%%%%%%%%
\begin{titlepage}
\begin{flushright}
March 1998
\end{flushright}
\vskip 1cm
\begin{center}
{\Large\bf Erratum: \\
\medskip
Jet Cluster Algorithms in $\ee$ Annihilation}
\vskip .8cm
{ \large S. Bethke}\\
{\it III. Physikalisches Institut,\\
RWTH Aachen, Germany \\}
\ \\
{ \large Z. Kunszt}\\
{\it Institute of Theoretical Physics,\\
ETH Z\"urich, Switzerland}\\
\ \\
{ \large D.E. Soper}\\
{\it Institute of Theoretical Science, University of Oregon, \\
Eugene, Oregon, USA}\\
\ \\
{ \large W.J. Stirling}\\
{\it Department of Physics, University of Durham, \\
  Durham, England}
\end{center}
\bigskip

\abstract{
We correct an important misprint in the journal version of our earlier work
on {\it "New Jet Cluster Algorithms: Next-to-leading Order QCD ..."},
published in Nucl. Phys. B 370 (1992) 310, which may
have lead to an incorrect parametrisation of the leading order QCD
coefficients for the JADE type jet cluster algorithms.
}
%\vskip4.0cm

%\begin{flushleft}
%Version of February 6, 1998 \\
%\end{flushleft}

\end{titlepage}
\newpage

Only recently we were made aware of a subtle misprint in the journal
version of our work on {\it "New Jet Cluster Algorithms: Next-to-leading
Order QCD and Hadronization Corrections"} \cite{bkss-np}; the preprint
version of that publication \cite{bkss-cern} is correct and does {\bf
not} contain this typing error.

The misprint in \cite{bkss-np} appears in the first line of numbers
presented in Table~2, where we gave a simple parametrisation of the QCD
coefficients for jet production rates in hadronic final states of $\ee$
annihilations. 
In particular, the parameter $k_1$ parametrising the leading order
coefficient $A(y)$ for the relative production rate of 3-jet events
using the JADE-type jet algorithms \cite{jadejet,opaljet}, 
$$f_3(y) =  A(y) {\as \over {2\pi}}\ ,$$ should correctly be -6.513 (as given
in
\cite{bkss-cern}) instead of -6.153 (as misprinted in \cite{bkss-np}).

This misprint, if implemented in experimental determinations of $\as$
from measurements of jet production rates using the JADE jet finding
algorithm, may result in an overestimate of $\as$ by 5 to 10~\%, depending
on the range of jet resolution parameters and centre of mass energy utilised
in the analysis.

The {\bf correct} part of Table~2 should therefore read

\begin{table}[h]
\begin{center}
\begin{tabular}{|l||l|c|r|r|r|r|r|}
   \hline
 & Algorithm & y range & $k_0$ & $k_1$ & $k_2$ & $k_3$ & $k_4$ \\
   \hline \hline
$A$ & E, E0, p & 0.01 - 0.33  &
         3.096 & {\bf -6.513} & 3.463 & -0.157 & 0.0134 \\
   \hline
   \end{tabular}
\end{center}
\end{table}

\noindent
where $A(y)$ is parametrised according to equation~22
of \cite{bkss-np,bkss-cern}:
$$
A(y) = \sum_{n=0}^{n=4} k_n \left( \log{1\over y} \right)^n \; .
$$
The coefficients $k_n$ for the leading- and the next-to-leading order
QCD predictions of the other jet schemes given in \cite{bkss-np,bkss-cern}
are correct and are not repeated here.

\medskip
\noindent
{\bf Acknowledgement} \\
We thank A. Brandenburg for making us aware of the misprint in Table~2 of
\cite{bkss-np}.

\end{document}